\documentclass[runningheads]{llncs}
\usepackage{graphicx}
\usepackage{makecell}
\usepackage{adjustbox}
\begin{document}
\title{Backporting RISC-V Vector assembly}
%
%
\author{Joseph K. L. Lee\inst{1}\orcidID{0000-0002-1648-2740} \and
Maurice Jamieson\inst{1}\orcidID{0000-0003-1626-4871} \and
Nick Brown\inst{1}\orcidID{0000-0003-2925-7275}}
\authorrunning{J. K. L. Lee et al.}
%
\institute{EPCC, University of Edinburgh, Bayes Centre, 47 Potterrow, Edinburgh, United Kingdom\\
\email{\{j.lee,m.jamieson,n.brown\}@epcc.ed.ac.uk}}
\maketitle              
\begin{abstract}
Leveraging vectorisation, the ability for a CPU to apply operations to multiple elements of data concurrently, is critical for high performance workloads. However, at the time of writing, commercially available physical RISC-V hardware that provides the RISC-V vector extension (RVV) only supports version 0.7.1, which is incompatible with the latest ratified version 1.0. The challenge is that upstream compiler toolchains, such as Clang, only target the ratified v1.0 and do not support the older v0.7.1. Because v1.0 is not compatible with v0.7.1, the only way to program vectorised code is to use a vendor-provided, older compiler. In this paper we introduce the rvv-rollback tool which translates assembly code generated by the compiler using vector extension v1.0 instructions to v0.7.1. We utilise this tool to compare vectorisation performance of the vendor-provided GNU 8.4 compiler (supports v0.7.1) against LLVM 15.0 (supports only v1.0), where we found that the LLVM compiler is capable of auto-vectorising more computational kernels, and delivers greater performance than GNU in most, but not all, cases. We also tested LLVM vectorisation with vector length agnostic and specific settings, and observed cases with significant difference in performance.

\end{abstract}

\section{Introduction}
Whilst the first proposal of the RISC-V vector extension (RVV) was introduced in June 2015, this was only ratified in late 2021. The goal of the vector extension is to be efficient and scalable, and the result is a Cray-style, variable sized vector model. RVV can reconfigure element size and vector length at run time, and is flexible so that it works on different data types such as integer, fixed-point and floating-point, and microarchitectures such as in-order, out-of-order and decoupled. When combined with the base ISA the total instruction count is around 300 instructions which is far fewer than typical packed-SIMD alternative, and fits into a standard fixed 32-bit encoded space \cite{vector_spec}. RVV also forms the foundation for other vector extensions, such as the vector cryptographic extension.

Prior to ratification at version 1.0, a draft version 0.7.1 was released in 2019. According to this release: \emph{version 0.7 is intended to be stable enough to begin developing toolchains, functional simulators, and initial implementations, though will continue to evolve with minor changes and updates}~\cite{risc-v-v-extension-v0p7}. With the warning that \emph{backwards-incompatible changes will be made prior to ratification}, toolchains, simulators, and hardware implementations were developed. 

The first, and currently only, mass-produced hardware implementation of the vector extension v0.7.1 is the T-Head XuanTie C906~\cite{noauthor_product_nodate}, which contains 128-bit wide vector registers and supports up to 32-bit vector elements. This is used in the low-cost, widely available Allwinner D1 SoC, which reuses their existing Arm SoC peripheral IP. As of yet, no commercially available hardware cores implementing v1.0 have been announced, only IP cores are available for soft-core designs. Since v0.7.1 was not ratified, upstream compilers and software do not, and will not, target this RVV version. 

The aim of this paper is to address the gap between v1.0, the target for current and future tool development, and v0.7.1, the version supported by available hardware. This paper is structured as follows, in Section \ref{sec:bg} we describe the background to this work by exploring the differences between v1.0 and v0.7.1 of RVV before surveying support in different toolchains and highlighting related work. Our \emph{rvv-rollback} tool is then presented in Section \ref{sec:tool} where we describe both the design and how this is to be leveraged within the compiler flow. Section \ref{sec:benchmark} then undertakes benchmarking comparisons between different compilers using our tool to better understand the performance properties of common toolchains and setting, before drawing conclusions and discussing further work in Section \ref{sec:conclusions}.

The key contributions of this paper are:
\begin{enumerate}
    \item We review the main differences between the ratified RVV v1.0 and implemented v0.7.1 by currently available hardware
    \item We present our \emph{rvv-rollback} tool designed for translating RVV v1.0 assembly code into v0.7.1
    \item We utilise our \emph{rvv-rollback} tool to test the auto-vectorisation of available compilers using the RAJA Performance Suite~\cite{noauthor_llnlrajaperf_2023} and explore the impact that settings and compilers have on the overarching performance obtained.
\end{enumerate}

\section{Background and related work}
\label{sec:bg}
\subsection{RVV version 1.0 vs version 0.7.1}
The RISC-V vector extension (RVV) adds 32 vector registers which are specified by two implementation-defined parameters, the maximum size in bits of a vector element, $ELEN \geq 8$, and the number of bits in a single vector register, $VLEN \leq 2^{16}$. RVV v0.7.1 adds five unprivileged Control and Status Registers (CSRs) \textit{vstart}, \textit{vxsat}, \textit{vxrm}, \textit{vl} and \textit{vtype}, whereas v1.0 extends this list with two additional registers, \textit{vcsr} (a vector control and status register) and \textit{vlenb} (vector register length in bytes). Among other information, these CSRs contain settings about the selected element width \emph{SEW}, vector register group multiplier \emph{LMUL} (the number of vector registers grouped together), and operational vector length \emph{vl} (the number of elements to be updated from a vector instruction).

Other important differences between RVV v1.0 and v0.7.1 include:
\begin{itemize}
\item \textbf{Configuration-setting instructions:} To update the vector type settings, configuration-setting instructions \emph{vsetvl} and \emph{vsetvli} have to be used, where the application specifies the element type and total number of elements to be processed. The hardware then configures the \emph{vl} and \emph{vtype} CSRs to match what is required by the application. RVV v1.0 introduced an extra instruction \emph{vsetivli}, where the application can provide an immediate value directly as the application vector length, enabling more compact code to be generated by the compiler.

\item \textbf{Fractional LMUL:} RVV allows multiple vector registers to be grouped together so that a single vector instruction can operate on multiple vector registers concurrently. This allows double-width, or larger, elements to be operated on with the same vector length as single-width elements. It is also possible for instructions to accept source and destination vector operands with differing element widths but the same number of elements, thus increasing flexibility. Vector register grouping can also improve the execution efficiency for longer application vectors because the hardware is then flexible enough to enable these to run concurrently.

The grouping is defined by the vector length multiplier \emph{LMUL} which represents the default number of vector registers that are combined together to form a vector register group. Implementations must support \emph{LMUL} of integer values 1, 2, 4, and 8. For v1.0, \emph{LMUL} can also accept the fractional values $\frac{1}{2}$, $\frac{1}{4}$ and $\frac{1}{8}$, which reduces the number of bits used in a single vector register. This is particularly useful when operating on mixed-width values, enabling the compiler to effectively increase the number of usable vector register groups.

\item \textbf{Tail/mask agnostic policy:} Tail elements are those which lie past the current vector length, \emph{vl}, setting. By contrast, inactive elements are those within the current vector length but are disabled by the current mask because they do not receive new results during a vector operation. For v0.7.1, all regular vector instructions place zeros in the tail elements of the destination vector register group, and inactive elements are undisturbed. For v1.0, these elements can be independently marked either undisturbed or agnostic. The agnostic setting allows for the corresponding destination elements to either retain their values or be overwritten with 1s, the pattern of which is not required to be deterministic when the instruction is executed with the same inputs. The agnostic policy was added in RVV v1.0 to increase efficiency when the inactive or tail values are not required for subsequent calculations. For v1.0 all configuration-setting instructions, \emph{vsetvl}, \emph{vsetvli} and \emph{vsetivli} must include the flags for whether it is following the tail and mask agnostic (\emph{ta} and \emph{ma}) or undisturbed policies (\emph{tu} and \emph{mu}).

\item \textbf{Other changes:} RVV v1.0 simplifies the mask register layout by mapping the mask bit for element $i$ to bit $i$ of the mask register. Furthermore, v1.0 also introduces several new instructions, such as \emph{vl1r} which is a whole register load instruction, and also renames some instructions for example the \emph{vfredsum.vs} has become \emph{vfredusum.vs}. It should also be noted that because instruction encodings are different between v1.0 and v0.7.1 they are not binary compatible.
\end{itemize}

\subsection{Toolchain support}
The current upstream RISC-V GNU compiler toolchain does not provide support for any version of the vector extension. Whilst the GNU repository does contain an \emph{rvv-next} branch \cite{rvv-next} which aims to support v1.0, at the time of writing this is not actively maintained. There is also a previous, and now deleted, \emph{rvv-0.7.1} branch which targeted v0.7.1. Because of this lack of GCC support, T-Head, who are the chip division of Alibaba, provides their own modified GNU compiler toolchain (XuanTie GCC) which has been optimised for their C906 processor. This bespoke compiler supports both RVV v0.7.1 and also their own custom extensions. Several versions of this compiler have been provided, and through experimentation we found that GCC8.4 as found in their 20210618 release (a mirror is available at the EPCC RISC-V testbed website~\cite{team_excalibur_nodate}) provides the best auto-vectorisation capability and-so is used for benchmark comparisons in Section \ref{sec:benchmark} because this version generates code specifically targeting 128-bit vector lengths.

By comparison, Clang 15, provided as part of LLVM supports RVV v1.0. Furthermore, programmers are able to target RVV assembly which is vector length agnostic via the flag \emph{scalable-vectorization=on} or vector length specific via the flag \emph{riscv-v-vector-bits-min=N} (where \emph{N} is the fixed vector width in bits). Therefore it can be stated that, at the time of writing, Clang provides greater support for RVV than vanilla GCC.

In this paper, we use the RAJA Performance Suite~\cite{noauthor_llnlrajaperf_2023} to test the auto-vectorisation performance across compilers for a range of loop-based computational kernels. We observed that T-Head's XuanTie GCC 8.4 is capable of vectorising fewer kernels than Clang 15 with either vector length agnostic (VLA) or vector length specific (VLS) settings. This can be seen in Table~\ref{tab:compilers-vec-list}, which lists the kernels which are able to be auto-vectorised by the different compilers at different settings. This demonstrates the benefit of being able to leverage Clang on existing RISC-V hardware, and furthermore as further developments to main branch versions of these compilers will only support RVV v1.0, in future vectorisation will only be usable on existing vector hardware if the code are translated to RVV v0.7.1. This is the aim of the \emph{rvv-rollback} tool we have developed and describe in this paper.

\subsection{Related work}

This paper aims to bridge the gap between compilers that are targeting RVV v1.0 and mass-produced physical hardware available for consumer purchase which only supports v0.7.1. \cite{perotti_new_2022} presents an upgrade of Ara, a vector co-processor design, and reviewed the differences between RVV v0.5 and v1.0 to study the design changes required for updating to v1.0. \cite{adit_performance_2022} studied the auto-vectorisation capability of Clang 15 for RVV via dynamic instruction counting, and identified areas of improvements including the requirement to undertake improved shuffle pattern analysis and outer-loop vectorisation. 

The RISC-V vector landscape and software ecosystem was surveyed in~\cite{jkll_riscv_vector_status_2023}, exploring results from benchmarks running on T-Head's C906 using the XuanTie GCC 8.4 compiler. However, the authors were unable to include Clang in their benchmarking because of the RVV version issues that we are looking to address in this paper. 

Vehave~\cite{noauthor_vehave_2021} developed by BSC is a runtime library that enables the execution of vector instructions conforming to RVV v0.7.1 on RISC-V CPUs which do not support the vector extension, effectively trapping the unknown instructions and emulating them in software. Whilst this approach provides the ability to simulate vector instructions, the trapping and execution in software is far slower than execution over hardware. By contrast, in our approach we directly modify the assembly code generated by the compiler to \emph{roll back} the vectorisation to the v0.7.1 standard, thus enabling the code to run on the hardware directly and not requiring any runtime support.

\section{The RVV rollback tool}
\label{sec:tool}
We have developed \emph{rvv-rollback}, a Python based tool that backports RVV v1.0 assembly code to v0.7.1 assembly, and this is available at~\cite{noauthor_riscvtestbedrvv-rollback_nodate}. Whilst this tool is capable of translating most v1.0 instructions into version v0.7.1, some lesser used features of v1.0 such as fractional LMUL are not yet supported. 

RVV v1.0 introduced instructions which are immediate value versions of those found in v0.7.1. This is where the RVV instruction being issued contains part of the data being operated upon, such as a constant, rather than loading this from a register. Examples are the configuration set instruction \emph{vsetivli}, and the whole register load/store \emph{vl1r} and \emph{vs1r}. By default, our tool will convert these instructions to first store the current vector configuration in memory, then reconfigure the vector settings, followed by performing the instruction itself, and finally restoring the setting from memory. However, this process adds some overhead and furthermore is often unnecessary because a temporary register can be used instead or the reconfigurations being issued by the compiler are simply redundant. In verbose mode, the tool will print out these instances and recommend alternative optimised configuration options, with the user then able to manually determine the appropriate translation. 

It should be noted that this tool is aimed primarily to aid benchmarking applications, and whilst we have tested it extensively we make no guarantee as to bit reproducibility between the Clang generated RVV v1.0 assembly and our translated v0.7.1 code.

\subsection{RVV rollback compiler workflow}
In order to use our tool and generate RVV v0.7 executables using Clang, the user follows the following steps:
\begin{enumerate}
    \item Compile with Clang to obtain RVV v1.0 assembly code with the appropriate vector flags, for instance \emph{-march=rv64gcv -O3 -mllvm --riscv-v-vector-bits-min=128} for VLS or \emph{-scalable-vectorization=on} for VLA. The \emph{-no-integrate-as} flag is also necessary as it directs the compiler to generate assembly which can be assembled by the GNU assembler in the third step. \footnote{the {\tt -save-temps} flag can be useful for saving all intermediate assembly files if this is desired by the programmer.}
    \item Translate the assembly code to RVV v0.7.1 using our \emph{rvv-rollback.py} Python tool
    \item Assemble the generated assembly code using T-Head's XuanTie GCC assembler, provided as part of v2.6.1 of the the Xuantie-900-gcc-linux toolchain (also available at~\cite{team_excalibur_nodate}). This is required because a RVV v0.7.1 conforming compiler is needed to translate the v0.7.1 assembly into machine code.
\end{enumerate}

It should be highlighted that those RAJA kernels which failed to automatically vectorise with T-Head's XuanTie GCC compiler detailed in Table \ref{tab:compilers-vec-list} are due to limitations in the front-end of the GCC compiler, where automatic vectorisation opportunities are identified and applied, rather than the assembler. Consequently, whilst we leverage the GNU assembler as our third step it does not reduce opportunities for automatic vectorisation that have been identified by Clang higher up in the compilation process.

\section{Benchmarking and Comparison}
\label{sec:benchmark}
To demonstrate the use of our RVV rollback tool and the compilation workflow described in Section \ref{sec:tool}, we utilise the RAJA Performance Suite compiled using Clang 15 (generating RVV v1.0 assembly and backported to v0.7.1 using \emph{rvv-rollback}) and XuanTie GCC 8.4 which natively generates an RVV v0.7.1 executable. The suite is compiled with single-precision floating point numbers (some double precision constants found within the code were manually converted to single precision). The compiler and relevant flags are listed in Table~\ref{table:compiler-specs}.

\begin{table}
\caption{Compiler specifications}\label{table:compiler-specs}
\centering
\begin{adjustbox}{width={\textwidth},totalheight={\textheight},keepaspectratio}
\begin{tabular}{|c|c|c|p{0.55\linewidth}|}
\hline
 Name & Compiler & RVV Version &Compiler flags \\
\hline \hline
 GCC8.4-scalar & XuanTie GCC 8.4 & N/A&{\tt -O3 -march=rv64gc -ffast-math} \\
 GCC8.4-vector & XuanTie GCC 8.4 & 0.7 & {\tt  -O3 -march=rv64gcv0p7 -ffast-math} \\
 Clang15-scalar & Clang 15.0 & N/A &{\tt --march=rv64gc -O3 -ffast-math}\\
 Clang15-vector-vls & Clang 15.0 & 1.0 & {\tt -march=rv64gcv -O3 -mllvm --riscv-v-vector-bits-min=128 -ffast-math}\\
 Clang15-vector-vla & Clang 15.0 & 1.0 &{\tt -march=rv64gcv -O3 -mllvm -scalable-vectorization=on -ffast-math}\\
\hline
\end{tabular}
\end{adjustbox}
\end{table}

In this section we compare vectorisation performance of these benchmarks across the compilers on the Allwinnner D1. For reference, we also include result from the popular StarFive VisionFive V2 board (VF2), which contains a non-vectorised StarFive JH7110 processor (quad core SiFive U74). For these results benchmarks are run on a single core to provide a like-for-like comparison. The details of the systems we use in our experiments are reported in Table~\ref{table:compute-system}. 

\begin{table}
\caption{Compute system specifications}\label{table:compute-system}
\centering
\begin{tabular}{|c|c|c|}
\hline
 & Allwinner D1 & StarFive JH7110 (VF2)\\
 \hline \hline
 Processor & XuanTie C906 & SiFive U74\\
 Processor clock speed & 1.0 GHz  & 1.5GHz\\
 Cores & 1  & 4\\
 Cache & \makecell{32 KB I-cache + \\32 KB D-cache}  & \makecell{32KB I-cache + \\32 KB D-cache \\ + 2MB L2} \\
 Memory & 512MB DDR3  & 8GB DDR4\\
 \hline 
 ISA & RV64GC+V0.7  & RV64GC\\
 Vector width & 128bit  & N/A\\
 \hline
\end{tabular}
\end{table}

\subsection{Performance results}

\begin{table}[]
\caption{List of RAJA Performance Suite kernels auto-vectorised by XuanTie GCC 8.4 and Clang 15.0 compilers. * denotes kernels vectorised by GCC 8.4 but only scalar code was executed during runtime, and $^\dag$ denotes kernels vectorised by Clang (VLS and VLA) but only scalar code was executed during runtime.}
\label{tab:compilers-vec-list}
\begin{adjustbox}{width={\textwidth},totalheight={\textheight},keepaspectratio}
\begin{tabular}{|>{\raggedright}p{0.6\linewidth}|c|c|c|}
\hline
                 \textbf{Kernels}                                                         & \makecell{\textbf{XuanTie} \\ \textbf{GCC8.4 vector}}             & \makecell{\textbf{Clang15} \\ \textbf{vector VLA}}        & \makecell{\textbf{Clang15} \\ \textbf{vector VLS}}        \\ \hline
\textbf{Algorithm:} MEMCPY, MEMSET, REDUCE\_SUM                                                    & \multirow{7}{*}{X} & \multirow{7}{*}{X} & \multirow{7}{*}{X} \\ 
\textbf{Apps:} ENERGY, FIR, PRESSURE                                                               &                    &                    &                    \\ 
\textbf{Basic:} SAXPY, SAXPY\_ATOMIC, REDUCE3\_INT                                                 &                    &                    &                    \\ 
\textbf{Lcals:} FIRST\_DIFF*, FIRST\_SUM*,   GEN\_LIN\_RECUR, HYDRO\_1D*, HYDRO\_2D*, TRIDIAG\_ELIM*               &                    &                    &                    \\ 
\textbf{Polybench:} 2MM$^\dag$, 3MM$^\dag$, ATAX, FDTD\_2D, GEMM$^\dag$,   GEMVER, GESUMMV, JACOBI\_1D*, JACOBI\_2D*, MVT &                    &                    &                    \\ 
\textbf{Stream:} ADD, COPY, DOT, MUL, TRIAD                                                        &                    &                    &                    \\ \cline{1-1}
\textbf{Total: 30  }                                                                               &                    &                    &                    \\ \hline
\textbf{Apps:} LTIMES, LTIMES\_NOVIEW, VOL3D                                                       & \multirow{5}{*}{}  & \multirow{5}{*}{X} & \multirow{5}{*}{X} \\ 
\textbf{Basic:} IF\_QUAD, INDEXLIST\_3LOOP,   INIT\_INIT\_VIEW1D, INIT\_VIEW1D\_OFFSET, INIT3, MAT\_MAT\_SHARED, MULADDSUB,   NESTED\_INIT, PI\_ATOMIC, PI\_REDUCE, REDUCE\_STRUCT, TRAP\_INT &
   &
   &
   \\ 
\textbf{Lcals: }DIFF\_PREDICT, EOS, INT\_PREDICT                                                   &                    &                    &                    \\ 
\textbf{Polybench:} FLOYD\_WARSHALL, HEAT\_3D                                                      &                    &                    &                    \\ \cline{1-1}
\textbf{Total: 21}                                                                                 &                    &                    &                    \\ \hline
\textbf{Algorithm:} SORT                                                                           & \multirow{4}{*}{}  & \multirow{4}{*}{}  & \multirow{4}{*}{X} \\ 
\textbf{Apps} CONVECTION3DPA, DEL\_DOT\_VEC\_2D,   DIFFUSION3DPA, HALOEXCHANGE\_FUSED, MASS3DPA, NODAL\_ACCUMULATION\_3D &
   &
   &
   \\ 
\textbf{Lcals:} PLANCKIAN                                                                          &                    &                    &                    \\ \cline{1-1}
\textbf{Total: 8}                                                                                  &                    &                    &                    \\ \hline
\textbf{Algorithm:} SCAN                                                                           & \multirow{6}{*}{}  & \multirow{6}{*}{}  & \multirow{6}{*}{}  \\ 
\textbf{Apps:} HALOEXCHANGE                                                                        &                    &                    &                    \\ 
\textbf{Basic:} INDEXLIST                                                                          &                    &                    &                    \\ 
\textbf{Lcals:} FIRST\_MIN                                                                         &                    &                    &                    \\
\textbf{Polybench:} ADI                                                                            &                    &                    &                    \\ \cline{1-1}
\textbf{Total: 5}                                                                                  &                    &                    &                    \\ \hline
\end{tabular}
\end{adjustbox}
\end{table}
Table~\ref{tab:compilers-vec-list} lists the kernels which are able to be auto-vectorised by the different compilers at different settings. As mentioned, T-Head's XuanTie GCC 8.4 is capable of vectorising fewer kernels than Clang 15 with either vector length agnostic (VLA) or vector length specific (VLS) settings. Out of the kernels listed in Table~\ref{tab:compilers-vec-list}, 22 were translated using the compiler workflow described in Section \ref{sec:tool}. For reporting performance comparisons in Figures ~\ref{fig:GCCNoVec}, \ref{fig:GCCVecNotUsed} and \ref{fig:GCCVec}, we group kernels into three separate categories:

\begin{enumerate}
    \item Those kernels not vectorised by T-Head's XuanTie GCC 8.4 compiler
    \item Kernels vectorised by XuanTie GCC 8.4, but the kernel executed scalar code instead of vectorised code
    \item Kernels vectorised by XuanTie GCC 8.4, and the vectorised code was executed
\end{enumerate}

Figures ~\ref{fig:GCCNoVec}, \ref{fig:GCCVecNotUsed} and \ref{fig:GCCVec} report the runtime for each kernel compiled using GCC 8.4 and Clang 15.0 with scalar and vector for the Allwinner D1, and scalar for VF2. All runtimes are averaged across three runs and normalised against scalar code compiled with GCC8.4 on the Allwinner D1. 
\begin{figure}
\caption{Runtime for RAJA Performance Suite kernels normalised against Allwinner D1 with GCC8.4 scalar}
\centering
     \begin{subfigure}[b]{\textwidth}
         \centering
         \includegraphics[width=\textwidth]{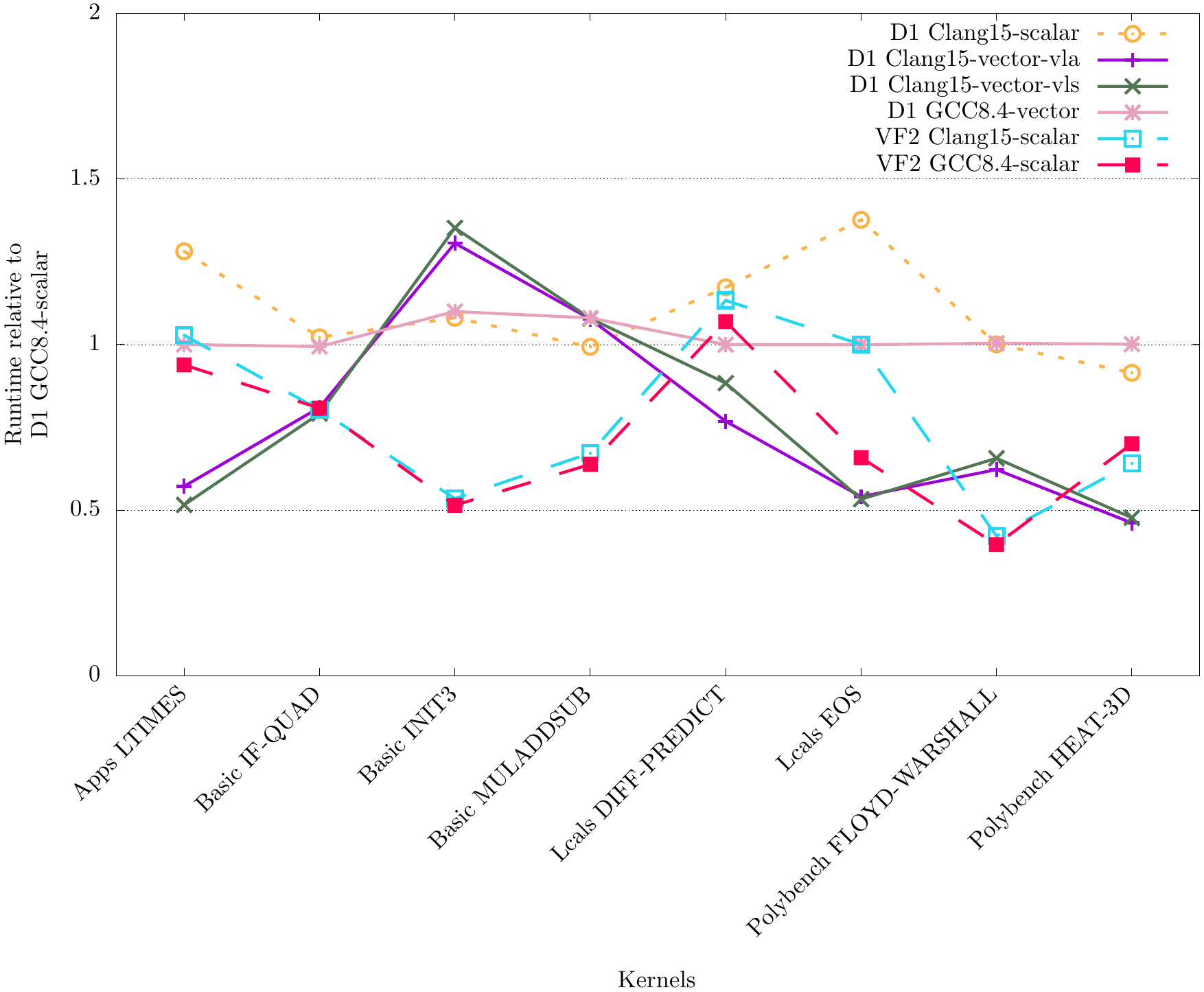}
         \caption{Kernels not vectorised by GCC8.4-vector}
         \label{fig:GCCNoVec}
     \end{subfigure}
      \begin{subfigure}[b]{\textwidth}
         \centering
         \includegraphics[width=\textwidth]{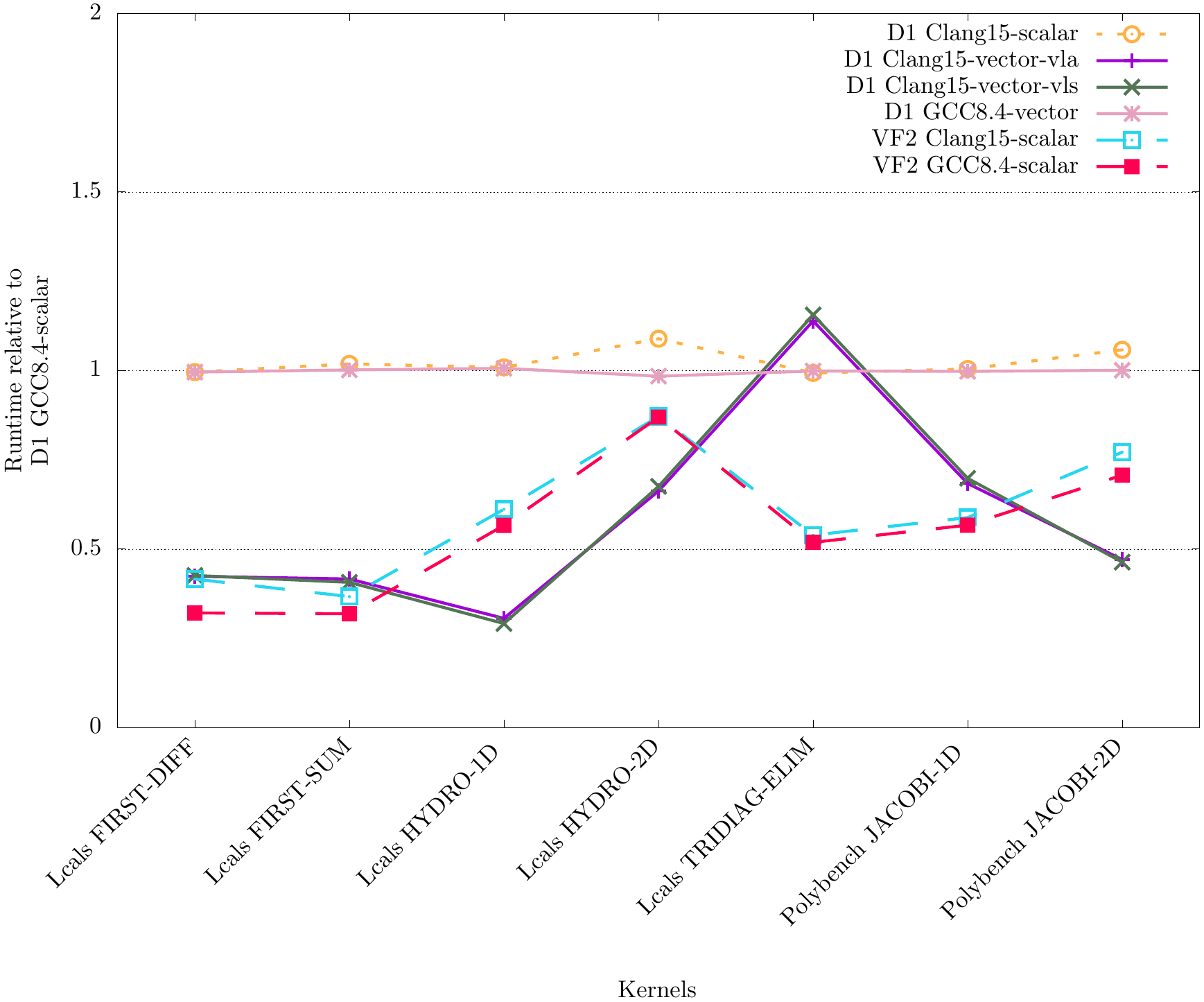}
         \caption{Kernels vectorised by GCC8.4-vector but only scalar code executed}
         \label{fig:GCCVecNotUsed}
     \end{subfigure}
       \end{figure}%
\begin{figure}[ht]\ContinuedFloat
     \begin{subfigure}[b]{\textwidth}
         \centering
         \includegraphics[width=\textwidth]{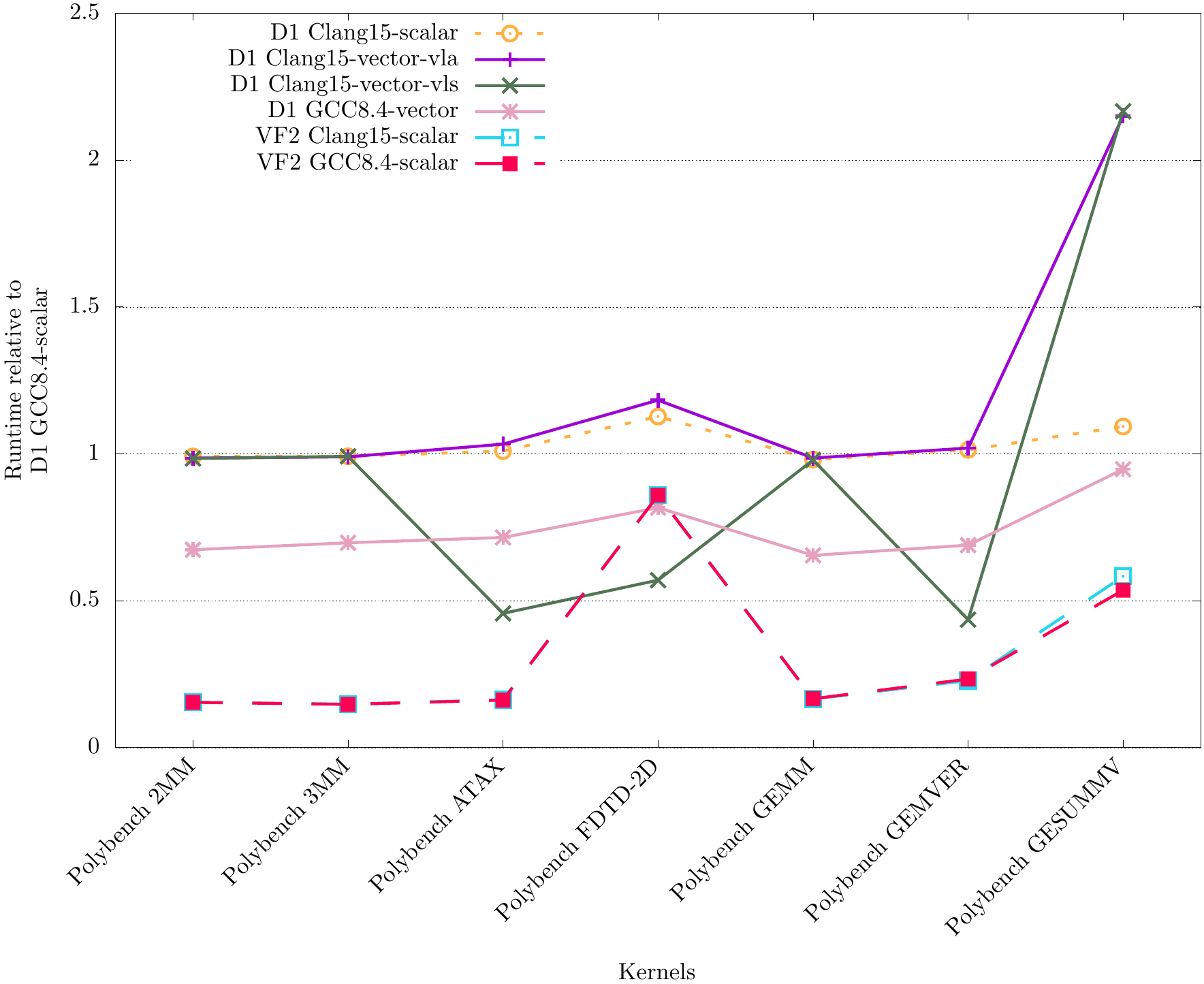}
         \caption{Kernels vectorised by GCC8.4-vector and vector code executed}
         \label{fig:GCCVec}
     \end{subfigure}
 \end{figure}
 
There are a number of noticeable features and behaviours that can be highlighted in these figures. Firstly it can be seen that Clang is capable of vectorising more kernels than GCC, especially for the LCALS routines, and this provides a significant speedup as seen when comparing Clang and GCC results in Figures~\ref{fig:GCCNoVec} and \ref{fig:GCCVecNotUsed}. However, for some kernels such as INIT3, TRIDIAG\_ELIM, and GESUMMV Clang's vectorised code is slower than its scalar counterpart.

For the 2MM, 3MM and GEMM matrix multiplication kernels whose performance is reported in Figure \ref{fig:GCCVec} it can be seen that Clang's vectorised performance exactly matches that of its scalar performance. This is because, whilst Clang was able to auto-vectorise the routines, the scalar code was executed, whereas by contrast for these benchmarks GCC executed its vectorised code and produces significantly faster runtimes. It can be seen therefore that whilst the auto-vectorisation reported in Table~\ref{tab:compilers-vec-list} demonstrates that on the whole Clang is able to vectorise more kernels than GCC, there are some exceptions to this rule.

Across most of the benchmark kernels Clang VLA (vector length agnostic) and VLS (vector length specific) settings provide very similar performance, except for specific kernels such as ATAX, FDTD\_2D and GEMVER. This demonstrates that it is important to experiment with these different compiler settings as it can make a difference in some situations to the achieved performance.

When comparing the performance of non-vectorised, scalar, code execution, it can be seen that for almost all kernels Clang 15 and GCC8.4 provide very similar performance, often within around 10\% of each other. However several kernels are an exception to this rule, for instance GCC is 52\% faster for the EOS kernel, 29\% faster for FIRST\_DIFF and 15\% faster for FIRST\_SUM. When comparing scalar performance on the Allwinner D1 against a single core of the U74 which is in the VisionFive V2, it can be observed that the V2 is significantly faster for high arithmetic intensity kernels, such as GEMM, compared to the Allwinner D1 running either vector or scalar code. However, for most kernels the vectorised kernels running on the Allwinner D1 is comparable with, if not faster than, the U74. This is especially impressive considering that the Allwinner D1 is considerably cheaper than the VisionFive V2, although it should be highlighted that we are comparing single-core performance here and unlike the D1 the U74 contains four compute cores so would likely deliver greater performance in practice.

A more general observation across our benchmark kernels was that we found when it comes to the compiler determining whether to generate vectorised or scalar instructions for execution depends heavily on loop ranges, which both compilers tend to be very sensitive to. For example, for some kernels the vectorised code is run only when the loop range is divisible by 8, and this demonstrates that it is therefore crucial that users manually check whether vectorised code is being emitted by the compiler, and executed, after compilation in order to obtain best performance.


\section{Conclusions, recommendations and future work}
\label{sec:conclusions}
In this paper we have explored compiler toolchains that enable vectorisation on mass-produced, commodity available RISC-V physical hardware. Whilst there is no main branch version of GCC that supports RISC-V vectorisation, a bespoke version by T-Head based on GCC 8.4 does support v0.7.1. However, as illustrated in Table~\ref{tab:compilers-vec-list}, it is less capable of automatic vectorisation compared to Clang 15. The challenge with Clang is that this only supports RVV v1.0 and-so we have introduced our tool, \emph{rvv-rollback}, to backport the generated v1.0 assembly to v0.7.1 

We have demonstrated that our tool runs across a wide set of benchmark codes, and the gathered performance numbers have illustrated that, in the main, vectorisation via Clang 15 is beneficial compared to T-Head's GCC 8.4 although there are always exceptions to this rule. Furthermore, we have demonstrated that whilst for most of our benchmark kernels the performance difference when compiling using VLA or VLS via Clang is narrow, for some codes it can make a more significant difference and-so it is important for programmers to experiment with these compiler flags.

One of the surprising aspects for us was that whilst the compiler will report that it has auto-vectorised code, it can sometimes revert to executing scalar-only code without the programmer knowing. Therefore it is crucial that programmers are aware of this and manually check what has been generated. One of our recommendations is that Clang should be clearer on this and also improve range checking to reduce the sensitivity around whether it picks one path or the other. Furthermore, effort should be invested into investigating why Clang is currently unable to execute auto-vectorised matrix multiplication operations.

In terms of future work, at the time of writing Clang 16 was released just a couple of days ago. Whilst we do not anticipate that this will have any impact on our \emph{rvv-rollback} tool, it will be interesting to explore whether the performance insights reported in Section \ref{sec:benchmark} have changed at all due to this latest version.

\section{Acknowledgement}
The authors would like to thank the ExCALIBUR H\&ES RISC-V testbed for access to compute resource and for funding this work.
%
%
%

%
%
%
 \bibliographystyle{splncs04}
 \bibliography{reference}
%
\end{document}